\documentclass[a4paper,11pt]{article}%
\usepackage{geometry}
\usepackage{amsmath}
\usepackage{amsfonts}
\usepackage{amssymb}
\usepackage{graphicx}
\usepackage{indentfirst}
\usepackage{bbold}
\usepackage[small,bf]{caption}
\usepackage{slashed}
\providecommand{\U}[1]{\protect\rule{.1in}{.1in}}
\DeclareMathOperator{\Tr}{Tr}

\begin{document}

\date{}
\title{\textbf{Implementing the Gribov-Zwanziger framework in $\mathcal{N}=1$ Super Yang--Mills in the Landau gauge}}
\author{\textbf{M.~A.~L.~Capri}\thanks{caprimarcio@gmail.com}\,\,,
%\textbf{D.~Dudal}$^{b}$\thanks{david.dudal@ugent.be}\,\,,
\textbf{D.~R.~Granado}\thanks{diegorochagrana@uerj.br}\,\,,
\textbf{M.~S.~Guimaraes}\thanks{msguimaraes@uerj.br}\,\,,
\textbf{I.~F.~Justo }\thanks{igorfjusto@gmail.com}\,\,,\\
\textbf{L.~F.~Palhares}\thanks{leticiapalhares@gmail.com}\,\,,
\textbf{S.~P.~Sorella}\thanks{silvio.sorella@gmail.com}\,\,\,,
\textbf{D.~Vercauteren}\thanks{vercauteren.uerj@gmail.com}\\[2mm]
{\small \textnormal{  \it Departamento de F\'{\i }sica Te\'{o}rica, Instituto de F\'{\i }sica, UERJ - Universidade do Estado do Rio de Janeiro,}}
 \\ \small \textnormal{ \it Rua S\~{a}o Francisco Xavier 524, 20550-013 Maracan\~{a}, Rio de Janeiro, Brasil}\normalsize}
	 %\small \textnormal{$^{b}$ \it Ghent University, Department of Physics and Astronomy, Krijgslaan 281-S9, 9000 Gent, Belgium}\normalsize}
%\author{\textbf{M.~A.~L.~Capri}$$\thanks{%
%primarcio@gmail.com}\,\,, \textbf{A.~J.~G\'{o}mez}\thanks{%
%ajgomez@uerj.br}\,\,, \textbf{M.~S.~Guimaraes}\thanks{%
%msguimaraes@uerj.br}\,\,, \textbf{S.~P.~Sorella}\thanks{%
%sorella@uerj.br}\, \thanks{%
%Work supported by FAPERJ, Funda{\c{c}}{\~{a}}o de Amparo {\`{a}} Pesquisa do
%Estado do Rio de Janeiro, under the program \textit{Cientista do Nosso Estado%
%}, E-26/101.578/2010.}\,\,,\,\,\\
%\textit{{\small { UERJ $-$ Universidade do Estado do Rio de Janeiro}}}%
%\\
%\textit{{\small {Instituto de F\'{\i}sica $-$ Departamento de F\'{\i}sica Te%
%\'{o}rica}}}\\
%\textit{{\small {Rua S{\~a}o Francisco Xavier 524, 20550-013 Maracan{\~a},
%Rio de Janeiro, RJ, Brasil.}}}}
\maketitle

\begin{abstract}
The Gribov-Zwanziger framework accounting for the existence of Gribov copies is extended to $\mathcal{N}=1$ Super Yang--Mills theories quantized in the Landau gauge. We show that the restriction of the domain of integration in the Euclidean functional  integral to the first Gribov horizon can be implemented in a way to recover non-perturbative features of $\mathcal{N}=1$ Super Yang--Mills theories, namely: the existence of the gluino condensate as well as the vanishing of the vacuum energy. 

\end{abstract}

\section{Introduction}
Supersymmetry is a powerful tool in order to investigate  non-perturbative aspects of gauge field theories. Supersymmetric gauge theories display remarkable non-renormalization features \cite{Grisaru_Siegel_Rocek:1979, Iliopoulos_Zumino:1974, Gates_Grisaru_Rocek_Siegel:1983, Sohnius_West:1981}, which follow from their holomorphicity properties \cite{Seiberg:1993, Seiberg:1988}.  As renewed examples, let us quote the work by  Seiberg and Witten \cite{Seiberg_Witten:1994} in ${\cal N}=2$ supersymmetric Yang-Mills theories,  in which the strong coupling regime has been described through the electromagnetic duality mechanism envisaged by 't Hooft and Mandelstam \cite{'tHooft:1981ht,Mandelstam:1974pi}. More recently, Maldacena's conjecture \cite{Maldacena:1997} has provided a framework to investigate the strong coupling regime of ${\cal N}=4$ Super-Yang-Mills, due to its duality with a weakly coupled supergravity theory in five dimensional anti-De Sitter space. \\\\Turning to the case of pure ${\cal N}=1$ Super-Yang-Mills theories, one notices similarity with QCD with one flavor of quarks, except for the fact that in Super-Yang-Mills gluinos are in the adjoint representation of the gauge group. Both theories display  confinement of their fundamental degrees of freedom, {\it i.e.} gluons and quarks in QCD and gauge field excitations and gluinos in ${\cal N}=1$ Super-Yang-Mills.  Though, while many aspects of the non-perturbative sector of QCD are still unknown, in the case of ${\cal N}=1$ SYM much progress has been done. For instance, holomorphicity has enabled  the computation of the gluino condensate \cite{Novikov_Shifman_Vainshtein_Zakharov:1983,Shifman_Vainshtein:1988}. Moreover, the exact beta function of theory has been evaluated \cite{Novikov_Shifman_Vainshtein_Zakharov:1982}. Let us also quote the  work by Veneziano and Yankielowicz \cite{Veneziano:1982ah} in which the low energy effective action describing the non-perturbative dynamics of the composite operators entering the so-called ${\cal N}=1$ anomaly supermultiplet has been derived. \\\\Concerning now non-supersymmetric gauge theories, much efforts have been done in the last decades in order to unravel their non-perturbative dynamics. Several aspects of the gluon and quark confinement have witnessed a better understanding as well as the corresponding formation of bound states, resulting in the observed mesonic and hadronic spectrum of QCD. Particular attention has been devoted to the  non-perturbative study of the Green  functions of asymptotically free Yang-Mills theories  in the infrared regime, which cannot be described by perturbation theory. Both numerical and analytic approaches have been employed. Let us mention, for instance,  the sophisticated numerical techniques employed in large lattice simulations to study the gluon and quark propagators and their connection with confinement  \cite{Cucchieri:2007rg,Cucchieri:2011ig,Oliveira:2012eh,Bornyakov:2008yx,Furui:2006ks,Parappilly:2005ei}. From the analytic side, let us quote the study of the Green functions through the Schwinger-Dyson and Bethe-Salpeter equations  \cite{Alkofer:2000wg,Binosi:2009qm} as well as the efforts to derive an effective Yang-Mills Lagrangian in the infrared region by taking into account the Gribov problem \cite{Gribov:1977wm}. \\\\As is widely known, the existence of the Gribov copies is  related to the gauge fixing procedure for quantizing Yang-Mills theories \cite{Singer:1978dk}, see \cite{Sobreiro:2005ec,Vandersickel:2012tz} for pedagogical reviews.  In recent years, it has become more and more clear that the issue of the Gribov copies is an important aspect of the non-perturbative dynamics of Yang-Mills theories with deep relation with confinement. Although a complete resolution of the Gribov issue is still lacking\footnote{See, for instance, the recent work \cite{Landim:2014fea} and refs. therein.}, important results have been obtained, resulting in the so-called Gribov-Zwanziger set up  \cite{Gribov:1977wm,Zwanziger:1988jt,Zwanziger:1989mf,Zwanziger:1992qr}, which allows one to take into account the existence of the Gribov copies in a local and renormalizable way. In this framework, the issue of the Gribov copies is faced by restricting the domain of integration in the Euclidean functional integral to the so-called Gribov region $\Omega$, which is bounded by the first Gribov horizon, where the first vanishing eigenvalue of the Faddeev-Popov operator shows up. Remarkably, this restriction results in a local and renormalizable action encoding non-perturbative aspects of the infrared dynamics of Yang-Mills theories. This action is known as the Gribov-Zwanziger action \cite{Gribov:1977wm,Zwanziger:1988jt,Zwanziger:1989mf,Zwanziger:1992qr}. More recently, a refinement of the Gribov-Zwanziger action has been worked out by the authors  \cite{Dudal:2007cw,Dudal:2008sp,Dudal:2011gd}, by taking into account the existence of certain dimension two condensates.  So far, the outcome obtained by employing the Refined-Gribov-Zwanziger (RGZ) action can be considered rather promising. Let us give here a short overview of what has been done. Let us start by mentioning that the gluon propagator resulting from the RGZ action exhibits complex poles and violates the reflection positivity. This kind of two-point function lacks the  K{\"a}ll{\'e}n-Lehmann spectral representation and cannot be associated with the propagation of physical particles. Rather, it indicates that, in the non-perturbative infrared region, gluons are not physical excitations of the spectrum of the theory, {\it i.e.} they are confined. Remarkably, the gluon propagator obtained from the RGZ  action turns out to be in excellent agreement with the most recent  numerical lattice simulations done on large lattices \cite{Cucchieri:2007rg,Cucchieri:2011ig}. Also, the  RGZ propagator has been employed in analytic calculation of the first glueball states  \cite{Dudal:2010cd,Dudal:2013wja}, yielding results which compare well with the available numerical simulations as well as with other approaches, see \cite{Mathieu:2008me} for a recent account on this topic. Recently, the RGZ propagator has been employed in the study of the evaluation of the Casimir energy within the MIT bag model \cite{Canfora:2013zna}. The resulting energy has the correct expected confining behaviour. Applications  of the RGZ gluon propagator at finite temperature can be found in  \cite{Fukushima:2013xsa,Canfora:2013kma}. Finally, in \cite{Capri:2012ah,Capri:2013oja}, the issue of the Gribov copies has been addressed in the case in which Higgs fields are present, yielding analytic  findings on  the hard problem of the understanding of the transition between the confining and Higgs phases for asymptotically free gauge theories. The output of this analysis  turns out to be in qualitative agreement with the seminal work by  Fradkin-Shenker \cite{Fradkin:1978dv}. All these results enable us to state that the issue of the Gribov copies  captures nontrivial aspects of the non-perturbative dynamics of Yang-Mills theories. \\\\It seems thus natural to extend the Gribov-Zwanziger framework to supersymmetric Yang-Mills theories in order to investigate their non-perturbative features. This is the goal of the present work. More precisely, we shall  extend the Gribov-Zwanziger framework to $\mathcal{N}=1$ Super-Yang-Mills theories quantized in the Wess-Zumino gauge with the Landau gauge-fixing condition.  There are several motivations in order to accomplish this task. The Gribov issue could provide an understanding from a different point of view of non-perturbative  supersymmetric features like the formation of the gluino condensate, the vanishing of the vacuum energy, the study of the states of the spectrum, etc. Moreover, similarly to the case of Yang-Mills theories, the supersymmetric extension of the Gribov-Zwanziger  could give us a  framework to investigate the confinement of the elementary degrees of freedom in $\mathcal{N}=1$ supersymmetric theories, {\it i.e.} gauge field excitations and gluinos, through the study of their two-point correlation functions. Let us also remind that, nowadays,  supersymmetric theories are object of  increasing investigation through numerical lattice simulations, see for instance \cite{Farchioni:2004fy} and refs. therein.  It is thus not unconceivable that, in the near future, the analytic predictions of the supersymmetric extension of the Gribov-Zwanziger framework could be compared with numerical data, just as in the case of Yang-Mills theories. \\\\Although in the present work we shall limit ourselves to discuss the case of $\mathcal{N}=1$ supersymmetric  theories, let us elaborate a little bit on the possible future perspectives that the present investigation could open. Certainly, it would be very interesting to extend the Gribov-Zwanziger framework to the case of $\mathcal{N}=2$ Super-Yang-Mills, especially in view of a  possible relation with the Seiberg-Witten work \cite{Seiberg_Witten:1994} on the confining phase of these theories. Also, the understanding of the Gribov issue in $\mathcal{N}=4$ supersymmetric theories could be of great relevance in order to study non-perturbative aspects of these theories within a pure quantum field theory framework and investigate a possible relationship with Maldacena's conjecture \cite{Maldacena:1997}. \\\\The present work is organized as follows. In Sect.2 we present the construction of the Gribov-Zwanziger framework for $\mathcal{N}=1$ Super-Yang-Mills theories. For the benefit of the reader, this section has been divided into several sub-sections. After giving a short summary of the Gribov-Zwanziger theory, we proceed with its $\mathcal{N}=1$ extension.   The resulting gauge field and gluino two-point functions are evaluated and we  analyze their connection with confinement.  In Sect.3 we discuss the formation of the gluino condensate within the $\mathcal{N}=1$ generalization of the Gribov-Zwanziger framework. In Sect.4 we address the issue of the vanishing of the vacuum energy. Sect.5 collects our conclusion.

%%%%%%%%%%%%%%%%%%%%%%%%%%%%%%%%%%%%%%%%%%%%%
\section{Extension of the Gribov-Zwanziger framework to $\mathcal N=1$ Super Yang--Mills}
%%%%%%%%%%%%%%%%%%%%%%%%%%%%%%%%%%%%%%%%%%%%%

\subsection{A brief summary of the Gribov-Zwanziger action in Yang-Mills theories} 

Let us start by giving a short overview of the Gribov-Zwanziger framework \cite{Gribov:1977wm,Zwanziger:1988jt,Zwanziger:1989mf,Zwanziger:1992qr}. As already mentioned in the Introduction, the Gribov-Zwanziger action arises from the restriction of the domain of integration in the Euclidean functional integral to the so-called Gribov region $\Omega$, which is defined as the set of all gauge  
field configurations fulfilling the Landau gauge, $\partial_\mu A^{a}_\mu=0$,  and for which the Faddeev-Popov operator ${\cal M}^{ab}=-(\partial^2 \delta^{ab} -g f^{abc}A^{c}_{\mu}\partial_{\mu})$ is strictly positive, namely 
\begin{align}
\Omega \;= \; \{ A^a_{\mu}\;; \;\; \partial_\mu A^a_{\mu}=0\;; \;\; {\cal M}^{ab}=-(\partial^2 \delta^{ab} -g f^{abc}A^{c}_{\mu}\partial_{\mu})\; >0 \; \} \;. \label{gr}
\end{align} 
One starts with the Faddeev-Popov action in the Landau gauge 
\begin{equation}
S_{FP} = S_{YM} + S_{gf}  \;, \label{fp}
\end{equation}
where $S_{YM}$  and $S_{gf}$ denote, respectively, the Yang-Mills and the gauge-fixing terms, namely 
\begin{equation} 
S_{YM} = \frac{1}{4} \int d^{4}x \; F^{a}_{\mu \nu}F^{a}_{\mu\nu} \;, \label{YM}
\end{equation}
and  
\begin{equation}
S_{gf} = \int d^{4}x \left(  b^{a}\partial_{\mu}A^{a}_{\mu}
+\bar{c}^{a} \partial_{\mu}D^{ab}_{\mu}c^{b}  \right) \;,  \label{gf}
\end{equation}
where $({\bar c}^a, c^a)$ stand for the Faddeev-Popov ghosts, $b^a$ is the Lagrange multiplier implementing the Landau gauge, $D^{ab}_\mu =( \delta^{ab}\partial_\mu + g f^{acb}A^{c}_{\mu})$ is the covariant derivative in the adjoint representation of $SU(N)$, and $F^{a}_{\mu\nu}$ denotes the field strength 
\begin{equation}
F^{a}_{\mu\nu} = \partial_{\mu}A^{a}_{\nu} - \partial_{\nu}A^{a}_{\mu} + gf^{abc}A^{b}_{\mu}A^{c}_{\nu}\;. \label{fstr}
\end{equation}
Following \cite{Gribov:1977wm,Zwanziger:1988jt,Zwanziger:1989mf,Zwanziger:1992qr}, the restriction of the domain of integration in the path integral is achieved by adding to the Faddeev-Popov action $S_{FP} $ an additional term $H(A)$, called the horizon term, given by the following non-local expression 
\begin{align}
H(A)  =  {g^{2}}\int d^{4}x\;d^{4}y\; f^{abc}A_{\mu}^{b}(x)\left[ {\cal M}^{-1}\right]^{ad} (x,y)f^{dec}A_{\mu}^{e}(y)   \;,  \label{hf1}
\end{align}
where ${\cal M}^{-1}$ stands for the inverse of the Faddeev-Popov operator. For the partition function one writes \cite{Gribov:1977wm,Zwanziger:1988jt,Zwanziger:1989mf,Zwanziger:1992qr}
\begin{equation}
 Z_{GZ} = \;
\int_\Omega {\cal D}A\;{\cal D}c\;{\cal D}\bar{c}\; {\cal D} b\; e^{-S_{FP}}  =   \int {\cal D}A\;{\cal D}c\;{\cal D}\bar{c} \; {\cal D} b \; e^{-(S_{FP}+\gamma^4 H(A) -V\gamma^4 4(N^2-1))} 
\;, \label{zww1}
\end{equation}
where $V$ is the Euclidean space-time volume. The parameter $\gamma$ has the dimension of a mass and is known as the Gribov parameter. It is not a free parameter of the theory. It is a dynamical quantity, being determined in a self-consistent way through a gap equation called the horizon condition \cite{Gribov:1977wm,Zwanziger:1988jt,Zwanziger:1989mf,Zwanziger:1992qr}, given by 
\begin{equation}
\left\langle H(A)   \right\rangle_{GZ} = 4V \left(  N^{2}-1\right) \;,   \label{hc1}
\end{equation}
where the notation $\left\langle H(A)  \right\rangle_{GZ}$ means that the vacuum expectation value of the horizon function $H(A)$ has to be evaluated with the measure defined in eq.\eqref{zww1}. An equivalent all-order proof of eq.\eqref{hc1} can be given within the original Gribov no-pole condition framework \cite{Gribov:1977wm}, by looking at the exact ghost propagator in an external gauge field \cite{Capri:2012wx}.   \\\\Although the horizon term  $H(A)$, eq.\eqref{hf1}, is non-local, it can be cast in local form by means of the introduction of a set of auxiliary fields $(\bar{\omega}_\mu^{ab}, \omega_\mu^{ab}, \bar{\varphi}_\mu^{ab},\varphi_\mu^{ab})$, where $(\bar{\varphi}_\mu^{ab},\varphi_\mu^{ab})$ are a pair of bosonic fields, while $(\bar{\omega}_\mu^{ab}, \omega_\mu^{ab})$ are anti-commuiting. It is not difficult to show that the partition function $Z_{GZ}$  in eq.\eqref{zww1} can be rewritten as \cite{Zwanziger:1988jt,Zwanziger:1989mf,Zwanziger:1992qr}
\begin{equation}
 Z_{GZ} = \;
\int {\cal D}A\;{\cal D}c\;{\cal D}\bar{c}\; {\cal D} b \; {\cal D}{\bar \omega}\; {\cal D} \omega\; {\cal D} {\bar \varphi} \;{\cal D} \varphi \; e^{-S_{GZ}} \;, \label{lzww1}
\end{equation}
where $S_{GZ}$ is given by the local expression 
\begin{equation} 
S_{GZ} = S_{YM} + S_{gf} + S_0+S_\gamma  \;, \label{sgz}
\end{equation}
with
\begin{equation}
S_0 =\int d^{4}x \left( {\bar \varphi}^{ac}_{\mu} (-\partial_\nu D^{ab}_{\nu} ) \varphi^{bc}_{\mu} - {\bar \omega}^{ac}_{\mu}  (-\partial_\nu D^{ab}_{\nu} ) \omega^{bc}_{\mu}  + gf^{amb} (\partial_\nu  {\bar \omega}^{ac}_{\mu} ) (D^{mp}_{\nu}c^p) \varphi^{bc}_{\mu}  \right) \;, \label{s0}
\end{equation}
and 
\begin{equation}
S_\gamma =\; \gamma^{2} \int d^{4}x \left( gf^{abc}A^{a}_{\mu}(\varphi^{bc}_{\mu} + {\bar \varphi}^{bc}_{\mu})\right)-4 \gamma^4V (N^2-1)\;. \label{hfl}
\end{equation} 
In the local formulation of the Gribov-Zwanziger action, the horizon condition \eqref{hc1} takes the simpler form 
\begin{equation}
 \frac{\partial \mathcal{E}_v}{\partial\gamma^2}=0\;,   \label{ggap}
\end{equation}
where $\mathcal{E}_{v}(\gamma)$ is the vacuum energy defined by:
\begin{equation}
 e^{-V\mathcal{E}_{v}}=\;Z_{GZ}\;  \label{vce} \;.
\end{equation}
The local action $S_{GZ}$ in eq.\eqref{sgz} is known as the Gribov-Zwanziger action. Remarkably, it has been shown to be renormalizable to all orders \cite{Zwanziger:1988jt,Zwanziger:1989mf,Zwanziger:1992qr,Dudal:2007cw,Dudal:2008sp,Dudal:2011gd}. This important property of the Gribov-Zwanziger action is a consequence of a set of Ward identities which follows from the existence of a soft breaking of the BRST invariance induced by the Gribov parameter $\gamma$. In fact, introducing the nilpotent BRST transformations 
\begin{eqnarray}
\label{brst1}
sA^{a}_{\mu} &=& - D^{ab}_{\mu}c^{b}\;,\nonumber \\
s c^{a} &=& \frac{1}{2}gf^{abc}c^{b}c^{c} \;, \nonumber \\
s{\bar c}^{a} &=& b^{a}\;, \qquad \; \; 
sb^{a} = 0 \;, \nonumber \\
s{\bar \omega}^{ab}_\mu & = & {\bar \varphi}^{ab}_\mu \;, \qquad  s {\bar \varphi}^{ab}_\mu =0\;, \nonumber \\
s { \varphi}^{ab}_\mu&  = & {\omega}^{ab}_\mu  \;, \qquad s {\omega}^{ab}_\mu = 0 \;, 
\end{eqnarray}
it is immediately checked that the Gribov-Zwanziger action exhibits a soft breaking of the BRST symmetry, as summarized by the equation
\begin{equation}
s S_{GZ} = \gamma^2 \Delta  \;, \label{brstbr}
\end{equation}
where 
\begin{equation}
\Delta = \int d^{4}x \left( - gf^{abc} (D_\mu^{am}c^m) (\varphi^{bc}_{\mu} + {\bar \varphi}^{bc}_{\mu})   + g f^{abc}A^a_\mu \omega^{bc}_\mu            \right)  \;. \label{brstb1}
\end{equation}
Notice that the breaking term $\Delta$ is of dimension two in the fields. As such, it is a soft breaking. The properties of the soft breaking of the $BRST$ symmetry of the Gribov-Zwanziger theory and its relation with confinement have been object of intensive investigation in recent years, see  \cite{Baulieu:2008fy,Dudal:2009xh,Sorella:2009vt,Sorella:2010it,Capri:2010hb,Dudal:2012sb,Reshetnyak:2013bga}. Here, it suffices to mention that the broken identity  \eqref{brstbr} is deeply connected with the restriction to the Gribov region $\Omega$. Moreover, eq.\eqref{brstbr} can be translated into a set of softly broken Slavnov-Taylor  identities which ensure the all order renormalizability of the action $S_{GZ}$. The presence of the soft breaking term $\Delta$ turns out to be necessary in order to have a gluon propagator with the desired confining properties. Though, a set of BRST  invariant composite operators whose correlation functions exhibit the K{\"a}ll{\'e}n-Lehmann spectral representation with positive spectral densities can be consistently introduced \cite{Baulieu:2009ha}. These correlation functions can be employed to obtain mass estimates on the spectrum of the glueballs \cite{Dudal:2010cd,Dudal:2013wja}. \\\\Let us conclude this brief review of the Gribov-Zwanziger action by noticing that the terms $S_{gf}$ and $S_{0}$ in expression \eqref{sgz} can be rewritten in the form  of a pure BRST variation, {\it i.e.}
\begin{equation} 
S_{gf} + S_{0} = s \int d^4x \left( {\bar c}^a \partial_\mu A^a_\mu + {\bar \omega}^{ac}_{\mu}  (-\partial_\nu D^{ab}_{\nu} ) \varphi^{bc}_{\mu}  \right) \;, \label{exbrst}
\end{equation}  
so that 
\begin{equation} 
S_{GZ} = S_{YM}  + s \int d^4x \left( {\bar c}^a \partial_\mu A^a_\mu + {\bar \omega}^{ac}_{\mu}  (-\partial_\nu D^{ab}_{\nu} ) \varphi^{bc}_{\mu}  \right) +  S_{\gamma}\;, \label{sgz1}
\end{equation}  
from which eq.\eqref{brstbr} becomes apparent. 

\subsection{ The $\mathcal N=1$ Super Yang--Mills and its quantization} 

Let us now proceed by giving a few details on the quantization of  $\mathcal N=1$ Euclidean Super Yang--Mills. We shall follow reference \cite{Capri:2014jqa}, where an all order proof of the renormalization of the theory  through the BRST symmetry has been given. \\\\In the Wess-Zumino gauge, the action of $\mathcal N=1$ Euclidean Super Yang--Mills is given by the expression\footnote{All conventions and notations about Euclidean $\mathcal{N}=1$ supersymmetry can be found in the Appendix of  \cite{Capri:2014jqa}.}  
\begin{equation}
\label{SYM}
S_{SYM}^{N=1} = \int d^{4}x \left( \frac{1}{4}F^{a}_{\mu \nu}F^{a}_{\mu\nu} + \frac{1}{2}\bar{\lambda}^{a\alpha}(\gamma_{\mu})_{\alpha\beta}D^{ab}_{\mu}\lambda^{b\beta}+\frac{1}{2}\mathfrak{D}^a\mathfrak{D}^a\right)\;,  
\end{equation}
where  the gluino field, $\lambda^{a\alpha}$, is the supersymmetric partner of the gauge field $A^{a}_\mu$. It is a Majorana spinor in the adjoint representation of the gauge group. The auxiliary field $\mathfrak{D}^{a}$ is needed for the off-shell closure of the supersymmetric algebra \cite{Capri:2014jqa}. Following \cite{Capri:2014jqa}, the  most powerful and efficient way to quantize the theory is that of  constructing a generalized BRST operator $Q$ which collects both gauge and supersymmetric  field transformations, namely
\begin{equation}
Q = s + \epsilon^{\alpha} \delta_{\alpha}   \;, \label{qbrst}
\end{equation}
where s is the usual BRST operator for gauge transformations and $\delta_{\alpha}$ are the generators of the supersymmetric  transformations. The parameter $\epsilon^{\alpha}$  is a constant Majorana spinor carrying ghost number 1. It can be identified with the ghost spinor parameter corresponding to the supersymmetry generators $\delta_{\alpha}$. The operator $Q$ acts on the fields of the theory in the following way
\begin{eqnarray}
\label{brstsusytransf}
QA^{a}_{\mu} &=& - D^{ab}_{\mu}c^{b} +\bar{\epsilon}^\alpha(\gamma_\mu)_{\alpha\beta}\lambda^{a\beta}\;,\nonumber \\
Q\lambda^{a\alpha} &=& gf^{abc}c^{b}\lambda^{c\alpha} - \frac{1}{2}(\sigma_{\mu\nu})^{~\alpha\beta}\epsilon_\beta F_{\mu\nu}^{a}+ (\gamma_{5})^{\alpha\beta}\epsilon_{\beta} \mathfrak{D}^a\;, \nonumber \\
Q\mathfrak{D}^a &=& gf^{abc}c^{b}\mathfrak{D}^c - \bar{\epsilon}^{\alpha}(\gamma_{\mu})_{\alpha \beta}D_{\mu}^{ab}(\gamma_{5})^{\beta\eta}\lambda^{b}_{\eta} \;, \nonumber\\
Qc^{a} &=& \frac{1}{2}gf^{abc}c^{b}c^{c} - \bar{\epsilon}^\alpha(\gamma_{\mu})_{\alpha\beta}\epsilon^\beta A^{a}_{\mu}\;, \nonumber \\
Q\bar{c}^{a} &=& b^{a}\;, \nonumber \\
Qb^{a} &=& \nabla\bar{c}^{a}  \;, 
\end{eqnarray}
and 
\begin{equation} 
Q S_{SYM}^{N=1} = 0 \;, \label{qinv}
\end{equation}
where we have introduced  the translation operator $\nabla$ 
\begin{equation}
\nabla :=\bar{\epsilon}^{\alpha}
(\gamma_{\mu})_{\alpha\beta}\epsilon^{\beta}\partial_{\mu}\,.
\end{equation}
It is easy to verify that the operator $Q$  enjoys the following important property
\begin{equation} 
Q^{2} = \nabla :=\bar{\epsilon}^{\alpha}
(\gamma_{\mu})_{\alpha\beta}\epsilon^{\beta}\partial_{\mu}\,, \label{q2}
\end{equation}
which enables us to quantize the theory in a BRST invariant way  \cite{Capri:2014jqa}.  Eq.\eqref{q2} implies that the operator $Q$ is in fact nilpotent when acting on integrated local polynomials in the fields.  Thanks to property \eqref{q2}, the introduction of the gauge fixing term can be done by following the standard BRST framework. Adopting the Landau gauge condition, $\partial_\mu A^{a}_{\mu}=0$, we have  \cite{Capri:2014jqa}
\begin{equation}
S_{gf}^{N=1} = Q\int d^{4}x \left( \bar{c}^{a}\partial_{\mu}A^{a}_{\mu} \right)\;,
\end{equation}
which, according to \eqref{brstsusytransf}, reads
\begin{equation}
S_{gf}^{N=1} = \int d^{4}x \left( \bar{c}^{a}\partial_{\mu}D^{ab}_{\mu}c^{b} + b^{a}\partial_{\mu}A^{a}_{\mu}   - \bar{c}^{a}\bar{\epsilon}^{\alpha}(\gamma_{\mu})_{\alpha\beta}\partial_{\mu}\lambda^{a\,\beta} \right)\;.   \label{gfn1}
\end{equation}
Expression \eqref{gfn1} is the supersymmetric generalization of the Landau gauge, as it can be inferred from the presence of the additional term $\bar{c}^{a}\bar{\epsilon}^{\alpha}(\gamma_{\mu})_{\alpha\beta}\partial_{\mu}\lambda^{a\,\beta}$ which contains  the supersymmetry ghost ${\epsilon}^{\alpha}$ as well as the gluino field $\lambda^{a\beta} $. \\\\Therefore, for the quantized BRST invariant $\mathcal{N}=1$ Super-Yang--Mills action in the Landau gauge we have
\begin{eqnarray}
S &= & S_{SYM}^{N=1}+S_{gf}^{N=1}\nonumber\\
&= &\!\int d^{4}x \left( \frac{1}{4}F^{a}_{\mu \nu}F^{a}_{\mu\nu} + \frac{1}{2}\bar{\lambda}^{a\,\alpha}(\gamma_{\mu})_{\alpha\beta}D^{ab}_{\mu}\lambda^{b\,\beta} + \frac{1}{2}\mathfrak{D}^{2}+ b^{a}\partial_{\mu}A^{a}_{\mu}
+\bar{c}^{a}\left[\partial_{\mu}D^{ab}_{\mu}c^{b}  - \bar{\epsilon}^{\alpha}(\gamma_{\mu})_{\alpha\beta}\partial_{\mu}\lambda^{a\,\beta} \right]\right) \;,  \nonumber \\
\label{qactn1}
\end{eqnarray}
and 
\begin{equation}
QS=0  \;. \label{sn1inv}
\end{equation}
As shown in \cite{Capri:2014jqa}, equation \eqref{sn1inv} can be translated into a set of Slavnov-Taylor identities, ensuring the all order renormalizability of $\mathcal{N}=1$ Super-Yang-Mills. \\\\It is interesting to notice that, even though the presence of the supersymmetry-ghost term in the action is crucial for writing down Slavnov-Taylor identities and proving the all order renormalizability, the extra bilinear interaction between the ghost field $\bar c$ and the gluino field $\lambda$, {\it i.e.} $  \bar{c}^{a} \bar{\epsilon}^{\alpha}(\gamma_{\mu})_{\alpha\beta}\partial_{\mu}\lambda^{a\,\beta} $, has  no impact on physical predictions of the theory, given by $n$-point functions with zero ghost number. This is clearly seen due to the fact that the supersymmetry ghost $\bar \epsilon$ has a positive (nonzero) ghost number that cannot be compensated by any other parameter of the theory in observable correlation functions. Formally, this condition is implemented in any correlation function with total ghost number zero to all orders in $\bar \epsilon$ ,due to the ghost-number Ward identity\footnote{The same argument holds for the Gribov-Zwanziger extended theory to be presented below, due to the presence of an extended ghost number Ward identity.} for the one-particle-irreducible generating functional $\Gamma = S + O(\hbar)$:
\begin{eqnarray}
\int d^4x \left( c\frac{\delta \Gamma}{\delta c}-\bar c\frac{\delta \Gamma}{\delta \bar c} \right) +\bar \epsilon \frac{\partial \Gamma}{\partial \bar \epsilon}=0\,.   \label{gwid}
\end{eqnarray}
For instance, acting on expression \eqref{gwid} with the test operator $\frac{ \delta^n}{\delta A^{a_1}_{\mu_1}(x_1) ..... \delta A^{a_n}_{\mu_n}(x_n)}$ and setting all fields equal zero at the end, one gets
\begin{equation}
\bar \epsilon \frac{\partial }{\partial \bar \epsilon} \; \langle A^{a_1}_{\mu_1}(x_1) .....  A^{a_n}_{\mu_n}(x_n) \rangle_{1PI} =0 \;, 
\end{equation}
stating that the 1PI $n$-point Green function $\langle A^{a_1}_{\mu_1}(x_1) .....  A^{a_n}_{\mu_n}(x_n) \rangle_{1PI}$ is independent from $\epsilon$. 

\subsection{Extension of the Gribov-Zwanziger framework to $\mathcal{N}=1$ Super-Yang-Mills theory}

We are now ready to discuss the generalization of the Gribov-Zwanziger set up to the  $\mathcal{N}=1$ Super-Yang-Mills theory. Owing to expression \eqref{sgz1}, for the generalization of the Gribov-Zwanziger action to $\mathcal{N}=1$ SYM theories we have found the following expression
\begin{equation} 
S_{SGZ}^{N=1} = S_{SYM}^{N=1}  + Q \int d^4x \left( {\bar c}^a \partial_\mu A^a_\mu + {\bar \omega}^{ac}_{\mu}  (-\partial_\nu D^{ab}_{\nu} ) \varphi^{bc}_{\mu}  \right) +  S_{\gamma} + S_{\lambda} \;, \label{sgzn1}
\end{equation}  
where:
\begin{itemize} 

\item $S_{SYM}^{N=1}$  is the N=1 Super-Yang-Mills action given in eq.\eqref{SYM}, \\[3mm]

\item the term $Q \int d^4x \left( {\bar c}^a \partial_\mu A^a_\mu + {\bar \omega}^{ac}_{\mu}  (-\partial_\nu D^{ab}_{\nu} ) \varphi^{bc}_{\mu}  \right) $ is the generalization to $\mathcal{N}=1$ of the corresponding expression  of eq.\eqref{sgz1}, {\it i.e.} $s \int d^4x \left( {\bar c}^a \partial_\mu A^a_\mu + {\bar \omega}^{ac}_{\mu}  (-\partial_\nu D^{ab}_{\nu} ) \varphi^{bc}_{\mu}  \right) $,  where the BRST operator $s$ has been replaced by the generalized operator $Q$. The $Q$-transformations of the auxiliary localizing fields $({\bar \varphi}^{ac}_{\mu},\varphi^{ac}_{\mu},  {\bar \omega}^{ac}_{\mu}, {\omega}^{ac}_{\mu})$ are given by 
\begin{eqnarray}
Q\varphi^{ac}_{\mu} &=& \omega^{ac}_{\mu} \nonumber \\
Q\omega^{ac}_{\mu} &=& \nabla\varphi^{ac}_{\mu} \nonumber \\
Q{\bar \omega}^{ac}_{\mu} &=& {\bar \varphi}^{ac}_{\mu} \nonumber \\
Q{\bar \varphi}^{ac}_{\mu} &=& \nabla{\bar \omega}^{ac}_{\mu}\;, 
\end{eqnarray}
so that  the property \eqref{q2} is preserved, {\it i.e.}
\begin{equation}
Q^2 = \nabla \;. 
\end{equation} 
\\[3mm]
\item $S_{\gamma}$ is the horizon term in its local form, eq.\eqref{hfl}, namely 
\begin{equation}
S_\gamma =\; \gamma^{2} \int d^{4}x \left( gf^{abc}A^{a}_{\mu}(\varphi^{bc}_{\mu} + {\bar \varphi}^{bc}_{\mu})\right)-4 \gamma^4V (N^2-1)\;. \label{hfln1}
\end{equation} 
As already mentioned in the previous subsection, this term follows from the restriction of the domain of integration to the Gribov region $\Omega$, needed in order to take into account the Gribov copies affecting the Landau gauge condition $\partial_\mu A^a_\mu=0$. 
It is worth underlining here that the expression of the Horizon function, eq.\eqref{hf1}, and of the related
gap equation, eq.\eqref{hc1}, remain the same in supersymmetric theories, since the exact ghost propagator in an external gauge field is left unmodified by the presence of the extra term $  \bar{c}^{a} \bar{\epsilon}^{\alpha}(\gamma_{\mu})_{\alpha\beta}\partial_{\mu}\lambda^{a\,\beta} $. Therefore, the all-order result
of \cite{Capri:2012wx} applies as well to the case of supersymmetric gauge theories.   
\\[3mm]
\item the term $S_\lambda$ is given by 
\begin{equation}
S_\lambda = - \frac{1}{2}M^3\int d^{4}x \left( \bar{\lambda}^{a\alpha}\frac{\delta_{\alpha\beta}}{\partial^{2}}\lambda^{a\beta}\right) \;, \label{sslambda}
 \end{equation} 
where, for the time being,  the massive constant $M$ is  a free parameter.  The action $S_\lambda$ can be seen as the supersymmetric counterpart of the term $S_\gamma$. The introduction of such a term can be easily justified by looking at the explicit expression of the Horizon term, eq.\eqref{hf1}, which, when expanded in powers of the gauge field, has the following form 
 \begin{equation}
\gamma^4 H(A)  =   -  N g^{2} \gamma^4 \int d^4x\; A^{a}_{\mu}\frac{1}{\partial^{2}}A^{a}_{\mu}  +\; {\rm higher \; order \; terms}  \label{hexp} \;,  
\end{equation}
from which one can appreciate the similarity with the term $S_\lambda$. To some extent, the action $S_\lambda$ is the simplest action in the gluino field which can be introduced. As we shall see in the following, the parameter $M$ will play an important role in order to recover important features of  $\mathcal{N}=1$ Super-Yang-Mills theories, namely: the existence of a non-vanishing gluino condensate  $\langle \bar\lambda \lambda \rangle \neq 0$ as well as the vanishing of the vacuum energy.  Let us also point out  that, although presented in non-local form, the action $S_\lambda$  can be easily localized by means of a set of auxiliary spinor fields. The local version of expression  \eqref{sslambda} is given by  
\begin{equation}
\label{actlocal}
S_{\lambda} = \int d^{4}x \left[ \hat{\zeta}^{a\alpha} \partial^{2} \zeta^{a}_{~\alpha} - \hat{\theta}^{a\alpha}\partial^{2} \theta^{a}_{~\alpha} - M^{3/2}(\bar{\lambda}^{a\alpha}\theta^{a}_{~\alpha} + \hat{\theta}^{a\alpha}\lambda^{a}_{~\alpha})\right]\;.
\end{equation}
Integrating out the auxiliary fields $(\hat{\zeta}^{a\alpha}, \zeta^{a\alpha}, \hat{\theta}^{a\alpha},\theta^{a\alpha})$ allows one to recover the expression  \eqref{sslambda}.  The localizing fields $(\hat{\zeta}^{a\alpha},\zeta^{a\alpha})$ are bosonic while $(\hat{\theta}^{a\alpha},\theta^{a\alpha}) $ are fermionic. They form doublets under $Q$ transformations, {\it i.e.} 
\begin{eqnarray}
\label{loctrans}
Q\hat{\theta}^{a}_{\alpha} &=& \hat{\zeta}^{a}_{\alpha} \;; \nonumber \\
Q\hat{\zeta}^{a}_{\alpha} &=& \nabla\hat{\theta}^{a}_{\alpha} \;; \nonumber \\
Q\zeta^{a}_{\alpha} &=& \theta^{a}_{\alpha} \;; \nonumber \\
Q\theta^{a}_{\alpha} &=& \nabla\zeta^{a}_{\alpha}\;, 
\end{eqnarray}
which are easily seen to preserve the property
\begin{equation}
Q^2= \nabla  \;. \label{nbs}
\end{equation}
Let us also observe that expression \eqref{actlocal} can be written as 
\begin{equation}
\label{actlocal1}
S_{\lambda} =    Q  \int d^{4}x  \;\hat{\theta}^{a\alpha} \partial^{2} \zeta^{a}_{~\alpha} 
- M^{3/2} \int d^4x \;(\bar{\lambda}^{a\alpha}\theta^{a}_{~\alpha} + \hat{\theta}^{a\alpha}\lambda^{a}_{~\alpha})\;.
\end{equation}
\end{itemize}
In summary, for the generalization of the Gribov-Zwanziger action to  $\mathcal{N}=1$ Super-Yang-Mills theory we obtain the following local action  
\begin{eqnarray}
S_{SGZ}^{N=1} = S_{SYM}^{N=1} \; & + &  \; Q \int d^4x \left( {\bar c}^a \partial_\mu A^a_\mu + {\bar \omega}^{ac}_{\mu}  (-\partial_\nu D^{ab}_{\nu} ) \varphi^{bc}_{\mu}  + \hat{\theta}^{a\alpha} \partial^{2} \zeta^{a}_{~\alpha}  \right)  \nonumber \\ 
& + &  \gamma^{2} \int d^{4}x \left( gf^{abc}A^{a}_{\mu}(\varphi^{bc}_{\mu} + {\bar \varphi}^{bc}_{\mu})\right)-4 \gamma^4V (N^2-1) - M^{3/2} \int d^4x\; (\bar{\lambda}^{a\alpha}\theta^{a}_{~\alpha} + \hat{\theta}^{a\alpha}\lambda^{a}_{~\alpha}) 
 \;. \nonumber \\
 \label{lcsgzn1}
\end{eqnarray} 
Similarly to the case of the Gribov-Zwanziger action, see eq.\eqref{brstbr}, the action $S_{SGZ}^{N=1}$ exhibits a soft breaking of the $Q$ symmetry, namely 
\begin{equation}
Q S_{SGZ}^{N=1} = \gamma^2 \Delta_{\gamma}^{N=1} + M^{3/2}   \Delta_{\lambda}^{N=1} \;,   \label{sbn1}
\end{equation}
where the soft breakings $( \Delta_{\gamma}^{N=1},  \Delta_{\lambda}^{N=1}) $ are given by 
\begin{eqnarray} 
\Delta_{\gamma}^{N=1} & = &  \int d^{4}x \left( gf^{abc} (- D_\mu^{am}c^m  + \bar{\epsilon}^{\alpha} (\gamma_\mu)_{\alpha\beta} \lambda^\beta ) (\varphi^{bc}_{\mu} + {\bar \varphi}^{bc}_{\mu})   + g f^{abc}A^a_\mu ( \omega^{bc}_\mu +\nabla {\bar \omega}^{bc}_\mu)             \right)  \;, \label{deltag} \\[3mm]
\Delta_{\lambda}^{N=1} & = &  - \int d^4x \; \left( {\hat \zeta}^{a \alpha} \lambda^{a}_\alpha -  {\hat \theta}^{a}_{ \alpha} \left( gf^{abc}c^{b}\lambda^{c\alpha} - \frac{1}{2}(\sigma_{\mu\nu})^{~\alpha\beta}\epsilon_\beta F_{\mu\nu}^{a}+ (\gamma_{5})^{\alpha\beta}\epsilon_{\beta} \mathfrak{D}^a \right) \;\; + {\rm c.c.} \; \right)  \label{deltal} \;. 
\end{eqnarray} 
It is worth to  point out that the action $S_{SGZ}^{N=1}$ displays the correct limiting behaviours: 
\begin{itemize} 
\item when the non-perturbative parameters $(\gamma, M)$ are removed, {\it i.e.} set to zero, expression \eqref{lcsgzn1}  reduces to $\mathcal{N}=1$ Super-Yang-Mills action. It is easy in fact to check that the $Q$-exact part of \eqref{lcsgzn1} depending on the localizing fields  $(\bar{\omega}_\mu^{ab}, \omega_\mu^{ab}, \bar{\varphi}_\mu^{ab},\varphi_\mu^{ab})$ and $(\hat{\zeta}^{a\alpha}, \zeta^{a\alpha}, \hat{\theta}^{a\alpha},\theta^{a\alpha})$ can be integrated out giving a unity, 

\item also, upon removal of the spinor fields ${\lambda^a_{\alpha}}$, expression \eqref{lcsgzn1} reduces precisely to the Gribov-Zwanziger action $S_{GZ}$, eq.\eqref{sgz1}, 

\item even if being out of the aim of the present work, let us mention that, as in the case of the Gribov-Zwanziger action \cite{Zwanziger:1988jt,Zwanziger:1989mf,Zwanziger:1992qr,Dudal:2007cw,Dudal:2008sp,Dudal:2011gd}, the soft breaking identity  \eqref{sbn1} can be converted into a set of generalized Slavnov-Taylor identities which ensure the all order renormalizability of the action $S_{SGZ}^{N=1}$ \cite{prep}. The details of the proof of the renormalizability will be reported in a separate work. Let us limit ourselves to notice that this important feature follows from the renormalizability of the Gribov-Zwanziger action, of the $\mathcal{N}=1$ Super-Yang-Mills theory \cite{Capri:2014jqa} and of actions of the type of  $S_\lambda$, as discussed  in \cite{Baulieu:2009xr}. 

\end{itemize}

\subsection{The gauge field and gluino propagators} 

Having identified the $\mathcal{N}=1$ supersymmetric generalization of the Gribov-Zwanziger action, eq.\eqref{lcsgzn1}, let us have  a look at the gauge field and gluino propagators. For the gauge field we have a Gribov type propagator, {\it i.e.}
\begin{equation}
\langle A^a_\mu(p) A^b_\nu (-p) \rangle = \delta^{ab} \left( \delta_{\mu\nu} - \frac{p_\mu p_\nu}{p^2} \right) \frac{p^2}{p^4+2Ng^2 \gamma^4}   \;. \label{gg}
\end{equation} 
One observes that, due to the presence of the Gribov parameter $\gamma$, the expression \eqref{gg} exhibits complex poles:
\begin{equation} 
\frac{p^2}{p^4+2Ng^2\gamma^4}  = \frac{1}{2} \left( \frac{1}{p^2-i\sqrt{2N} g\gamma^2} + \frac{1}{p^2+i\sqrt{2N} g\gamma^2} \right) \;. \label{ig}
\end{equation} 
As such, the correlation function \eqref{gg} cannot be associated to the propagation of a physical particle. Rather, this feature is taken as evidence of the fact that  the elementary gauge field excitations described by the action $S_{SGZ}^{N=1}$, eq.\eqref{lcsgzn1}, are in fact confined  \cite{Zwanziger:1988jt,Zwanziger:1989mf,Zwanziger:1992qr,Dudal:2007cw,Dudal:2008sp,Dudal:2011gd,Baulieu:2009ha}. \\\\In order to evaluate the propagator of the gluino,  consider the quadratic terms in the gluino fields of the action $S_{SGZ}^{N=1}$, {\it i.e.} 
\begin{equation}
S^\text{quad}_{\lambda} = \int d^{4}x \left( \frac{1}{2}\bar{\lambda}^{a\alpha}(x)\left(\slashed{\partial}_{\alpha\beta} - \frac{M^{3}\delta_{\alpha\beta}}{\partial^{2}}\right) \lambda^{a\beta}(x) \right)\;,    \label{lqu}
\end{equation}
where the auxiliary fields $(\hat{\zeta}^{a\alpha}, \zeta^{a\alpha}, \hat{\theta}^{a\alpha},\theta^{a\alpha})$  have been already integrated out. From expression \eqref{lqu}, the two-point gluino correlation function is found  to be 
\begin{equation}
\langle \bar{\lambda}^{a}_{\alpha}(p) \lambda^{b}_{\beta}(-p) \rangle = \delta^{ab} \;\frac{\big( ip_{\mu}(\gamma_{\mu})_{\alpha\beta} + m(p^2)\delta_{\alpha\beta}\big)}{p^{2} + m^{2}(p^2)}\;,  \label{lprop}
\end{equation}
where the momentum dependent form factor $m(p^2)$ is given by 
\begin{equation} 
m(p^2) = \frac{M^{3}}{p^{2}}   \;.  \label{mp}
\end{equation}
Therefore, for the gluino propagator we get 
\begin{equation}
\langle \bar{\lambda}^{a}_{\alpha}(p) \lambda^{b}_{\beta}(-p) \rangle = \delta^{ab} \;\frac{\big( i p^4 p_{\mu}(\gamma_{\mu})_{\alpha\beta}  + p^2 M^3  \delta_{\alpha\beta}\big)}{p^{6} +M^6}\;.  \label{lqu1}
\end{equation} 
Again, one remarks the presence of complex poles in expression \eqref{lqu1}. Fermion propagators of the kind of  \eqref{lqu1} are frequently employed in the analysis of the chiral symmetry breaking in QCD, as they model in a good way effects of quark confinement see, for example, ref.\cite{Alkofer:2003jj}. One also notices that, due to the presence of the parameter $M$, the infrared behaviour of expression \eqref{lqu1} is deeply different from the behaviour of the free spinor propagator, $\frac{i \gamma_\mu p_\mu}{p^2}$, which is, however, recovered in the deep ultraviolet limit  
\begin{equation}
\langle \bar{\lambda}^{a}_{\alpha}(p) \lambda^{b}_{\beta}(-p) \rangle \Big|_{p\rightarrow \infty}  \sim  \delta^{ab} \;\frac{ i p_\mu \gamma_\mu}{p^2}  
\end{equation}
Let us conclude this section by noticing that, since we are dealing with Majorana fermions, there is no charge conservation. As such, in addition to the correlation function  $\langle \bar{\lambda}^{a}_{\alpha}(p) \lambda^{b}_{\beta}(-p) \rangle$,  we also have the propagators $\langle \lambda(p)\lambda(-p)\rangle$ and $\langle \bar{\lambda}(p)\bar{\lambda}(-p)\rangle$, given by 
\begin{equation}
\langle \lambda^{a\rho}(p)\lambda^{b}_{\beta}(-p)\rangle = - \frac{\big( ip_{\mu}(\gamma_{\mu})_{\alpha\beta} + m(p^2)\delta_{\alpha\beta}\big)\delta^{ab}C^{\alpha\rho}}{p^{2} + m^{2}(p^2)}  \;, 
\end{equation}
and
\begin{equation}
\langle \bar{\lambda}^{a}_{\alpha}(p)\bar{\lambda}^{b\tau}(-p)\rangle = \frac{\big( ip_{\mu}(\gamma_{\mu})_{\alpha\beta} + m(p^2)\delta_{\alpha\beta}\big)\delta^{ab}C^{\beta\tau}}{p^{2} + m^{2}(p^2)}\;,
\end{equation}
where $C^{\alpha\beta}$ is the charge conjugation matrix\footnote{For definitions and notations we remind to Appendix A of  \cite{Capri:2014jqa}.}.

%%%%%%%%%%%%%%%%%%%%%%%%%%%%
\subsection{The gluino condensate $\langle \bar\lambda \lambda \rangle_{cond}$ }
%%%%%%%%%%%%%%%%%%%%%%%%%%%%

Having at our disposal the expressions of the propagators, we can employ them in order to get a first estimate of the gluino condensate $\langle \bar\lambda \lambda \rangle_{cond}$. This will provide us  with a better understanding of the role played by the parameter $M$. The gluino condensate $\langle \bar\lambda \lambda \rangle_{cond}$ is obtained by taking the trace of the two-point correlation function $\langle \bar\lambda^a_\alpha(x)  \lambda^b_\beta(y)  \rangle_{cond}$ at the same space-time point, {\it i.e.}  $\langle \bar\lambda \lambda \rangle_{cond}= {\rm lim}_{x\rightarrow y} Tr  \langle \bar\lambda^a_\alpha(x)  \lambda^b_\beta(y)  \rangle$, where the trace is taken over both color and Lorentz indices $(a,b)$ and $(\alpha, \beta)$. From the expression of the gluino propagator, eq.\eqref{lprop}, we get 
\begin{equation}
\langle \bar\lambda \lambda \rangle := \lim_{x \to y} \int \frac{d^{4}p}{(2\pi)^{4}} \Tr\langle \bar{\lambda}^{a}_{\alpha}(p) \lambda^{b}_{\beta}(-p) \rangle e^{ip.(x-y)} = 4(N^2-1)  \int \frac{d^{4}p}{(2\pi)^{4}}  \frac{m(p^2)}{p^2+m^2(p^2)}  \;. \label{gcond}
\end{equation}
Consistently with the known properties of supersymmetric $\mathcal{N}=1$ gauge theories, one immediately checks that, due to the property that the charge conjugation matrix $C^{\alpha\beta}$  is traceless, the condensate $\langle \lambda \lambda  \rangle_{cond}$ and $\langle \bar\lambda \bar\lambda \rangle_{cond}$ are absent, {\it i.e.}
\begin{equation}
\langle \lambda \lambda  \rangle_{cond} = \langle \bar\lambda \bar\lambda \rangle_{cond} = 0 \;. \label{vcond}
\end{equation}
Expression \eqref{gcond} shows in a direct way the role played by the mass form factor $m(p^2)$, eq.\eqref{mp}, and of its deep connection with the gluino condensate. Interestingly, a similar expression is found in QCD for the quark condensate $\langle q {\bar q} \rangle_{cond}$  \cite{Furui:2006ks,Parappilly:2005ei}. Substituting expression \eqref{mp} into eq.\eqref{gcond}, one gets 
\begin{equation}
\langle \bar\lambda \lambda \rangle_{cond}  = 4 (N^2-1) M^3 \int \frac{d^{4}p}{(2\pi)^{4}} \frac{ p^{2} }{p^{6} +  M^{6}}\;, \label{ggcc}
\end{equation}
which can be written as
\begin{equation}
\langle \bar\lambda \lambda \rangle_{cond} = 4(N^{2}-1)M^{3} \int \frac{d^{4}p}{(2\pi)^{4}} \sum_{i=1}^{3} \frac{\alpha_{i}}{p^{2} + m_{i}^{2}}\;,
\end{equation}
where $m_{i}$ are the three cubic roots of the denominator in the integrand of expression \eqref{ggcc},  and
\begin{subequations} \begin{gather}
\alpha_1 = \frac{-m_1^2}{(m_2^2-m_1^2)(m_3^2-m_1^2)} \;, \\
\alpha_2 = \frac{-m_2^2}{(m_1^2-m_2^2)(m_3^2-m_2^2)} \;, \\
\alpha_3 = \frac{-m_3^2}{(m_1^2-m_3^2)(m_2^2-m_3^2)} \;.
\end{gather} \end{subequations}
Mark that $\sum_i \alpha_im_i^2 = -1$, as can be easily verified by working it out.  
\noindent Making use of the $\overline{\text{MS}}$ renormalization scheme in $d=4-\varepsilon$ and of the standard integrals
\begin{equation}
\int \frac{d^{d}p}{(2\pi)^{d}} \frac{\alpha_{i}}{p^{2}+m_{i}^{2}} = \frac{\alpha_{i}m_{i}^{2}}{16\pi^{2}}\left(\ln\frac{m_{i}^{2}}{\bar{\mu}^{2}} - 1 \right) - \frac{\alpha_{i}m_{i}^{2}}{16\pi^{2}}\frac{2}{\varepsilon}  \;,
\end{equation}
one gets for the gluino condensate:
\begin{equation}
\langle \bar\lambda \lambda \rangle_{cond}  = 4(N^{2}-1) M^{3}\sum_{i=1}^{3} \frac{\alpha_im_i^2}{(4\pi)^2} \left(\ln\frac{m_i^2}{\bar\mu^2} - 1 \right)\;.
\label{gluinocondM}
\end{equation}
We see thus that a non-vanishing gluino condensate is obtained as far as the parameter $M$ is non-vanishing. This is the issue which will be faced in the next section, in which the requirement of the vanishing of the vacuum energy is employed as a powerful criterium in order to determine $M$.

%%%%%%%%%%%%%%%%%%%%%%%
\section{Vanishing of the vacuum energy}
%%%%%%%%%%%%%%%%%%%%%%%

Let us proceed by addressing another important feature of $\mathcal{N}=1$ supersymmetric gauge theories, namely the vanishing of the vacuum energy,  {\it i.e.} $\mathcal{E}_{v}=0$. It is a well-known  property of $\mathcal{N}=1$ supersymmetric gauge theories that they 
do exhibit a vanishing vacuum energy even in presence of a non-vanishing gluino condensate $\langle \bar\lambda \lambda \rangle_{cond}$ \cite{Veneziano:1982ah}.
Such an important feature will also be reproduced in the current extension of the Gribov-Zwanziger framework to $\mathcal{N}=1$ Super-Yang-Mills theories. 
The requirement of  a vanishing vacuum energy shall actually play an important role in our construction, as it will provide us a practical way of determining the parameter $M$ in a non-perturbative  fashion. 

In what follows we derive the nonperturbative vacuum energy of the $\mathcal{N}=1$ Super-Yang-Mills theories
in the Gribov-Zwanziger framework and show the fulfillment of the zero vacuum-energy condition in this context.
The procedure follows three steps: $(i)$ a perturbative computation, $(ii)$ the imposition of the Gribov  gap equation and $(iii)$ the zero vacuum energy condition. First, the vacuum energy is computed as usual from the zero-field, zero-source limit of the one-particle-irreducible generating functional $\Gamma^{N=1}_{SGZ}$ associated to the action $S^{N=1}_{SGZ}$, eq.\eqref{lcsgzn1}:
\begin{eqnarray}
\mathcal{E}_v(\gamma,M,\bar\mu,g)= \left.\Gamma_{SGZ}\right|_{{\rm fields}=0}\,,
\end{eqnarray}
being thus, in general, a function of\footnote{As argued above, due to the ghost number Ward identity, the vacuum energy is a physical observable with vanishing total ghost number and is therefore independent of the supersymmetric ghost $\bar\epsilon$ to all orders.} the Gribov parameter $\gamma$, its supersymmetric counterpart $M$, the renormalization scale $\bar\mu$, and the gauge coupling $g$.
The nonperturbative character of the result will then be introduced by the Gribov gap equation,
\begin{eqnarray}
\frac{\partial \mathcal{E}_v(\gamma,M,\bar\mu,g)}{\partial\gamma^2}=0\, ,
\end{eqnarray}
which fixes the Gribov parameter as $\gamma=\gamma^*(M,\bar\mu,g)\propto {\rm e}^{-\text{const}/g^2}$. Taking this solution for the Gribov parameter back into the expression of the vacuum energy, one arrives at
 the nonperturbative result $\mathcal{E}_v(\gamma^*(M,\bar\mu,g),M,\bar\mu,g)$. Finally, the vanishing of the vacuum energy of  $\mathcal{N}=1$ supersymmetric Yang-Mills theories becomes then a condition for fixing $M= M^*(\bar\mu,g)$, such that 
\begin{eqnarray}
\mathcal{E}_v\Big(\gamma^*\big(M^*(\bar\mu,g),\bar\mu\big),M^*(\bar\mu,g),\bar\mu,g \Big)=0\, .  \label{vc}
\end{eqnarray}
Having described the whole procedure, let us now turn to the actual evaluation of the vacuum energy at leading order. For this, we consider  the quadratic terms of the action, eq.\eqref{lcsgzn1}
\begin{eqnarray}
S_\text{quad} = \int d^{4}x \left[ \frac{1}{2} \left( \partial_{\mu}A^{a}_{\nu}\partial_{\mu}A^{a}_{\nu} - \left(1-\frac{1}{\alpha}\right)\partial_{\mu}A^{a}_{\nu}\partial_{\nu}A^{a}_{\mu}\right) + \frac{1}{2} \bar{\lambda}^{a\alpha}\left(\slashed{\partial}_{\alpha\beta} - \frac{M^{3}\delta_{\alpha\beta}}{\partial^{2}}\right)\lambda^{a\beta}\right. \nonumber \\
+ \left.
\bar{c}^a(p) \left(p^2\delta^{ab}\right)c^b(-p)
- N g^{2}\gamma^4A^{a}_{\mu}\frac{1}{\partial^{2}}A^{a}_{\mu}\right] -\gamma^4V4(N^2-1)\;,
\end{eqnarray}
where we have already omitted the  term $\sim \bar\epsilon$, which does not affect the result for the vacuum energy, and the limit $\alpha\to 0$ is implied in order to recover the Landau gauge condition.
In  Fourier space one gets
\begin{eqnarray}
S_\text{quad} &=& \frac{1}{2} \int \frac{d^{4}p}{(2\pi)^{4}}\Big[ A^{a}_{\mu}(p)\left( 
p^2\delta_{\mu\nu}-\left(1-\frac{1}{\alpha}\right)p_\mu p_\nu
+ 2N g^{2}\gamma^4\frac{\delta_{\mu\nu}}{p^{2}}\right) A^{a}_{\nu}(-p) 
\nonumber\\
&&+
\bar{c}^a(p) \left(p^2\delta^{ab}\right)c^b(-p)
+ \bar{\lambda}^{a\alpha}(p) \left(-i\slashed{p}_{\alpha\beta} + \frac{M^{3}\delta_{\alpha\beta}}{p^{2}}\right)\lambda^{a\beta}(-p)\Big]
-\gamma^4V4(N^2-1)\;.
\end{eqnarray}
The vacuum energy is then related to the partition function as $\mathcal{E}_v=-(1/V)\ln Z_{\text{quad}}$, with $Z_\text{quad}=\int \mathcal{D}[\text{fields}]\,{\rm e}^{-S_{\text{quad}}}$. For the partition function in the quadratic approximation, we have 
\begin{eqnarray}
Z_\text{quad} = \int \mathcal{D}A \mathcal{D}c \mathcal{D}\bar c \mathcal{D}\lambda \exp\Big\{ - \frac{1}{2}\int \frac{d^{4}p}{(2\pi)^{4}} \Big[ A^{a}_{\mu}(p)\mathcal{P}_{\mu\nu}^{ab}A^{b}_{\nu}(-p) +
\bar{c}^a(p) \left(p^2\delta^{ab}\right)c^b(-p)
 \nonumber\\+
 \bar{\lambda}^{a\alpha}(p)\mathcal{Q}_{\alpha\beta}^{ab}\lambda^{b\beta}(-p)\Big] 
 +\gamma^4V4(N^2-1)\Big\}\;,
% \nonumber\\
\end{eqnarray}
where
\begin{equation}
\mathcal{P}_{\mu\nu}^{ab} = \left(
p^2\delta_{\mu\nu}-\left(1-\frac{1}{\alpha}\right)p_\mu p_\nu
+ 2N g^{2}\gamma^4\frac{\delta_{\mu\nu}}{p^{2}}
\right)
\delta^{ab}
\end{equation}
and
\begin{equation}
\mathcal{Q}_{\alpha\beta}^{ab} = \left( -i\slashed{p}_{\alpha\beta} + \frac{M^{3}\delta_{\alpha\beta}}{p^{2}}\right)\delta^{ab}\;.
\end{equation}
Integrating over the fields one gets
\begin{eqnarray}
\label{z1}
Z_\text{quad} &=&  \left[ \det \mathcal{P}^{ab}_{\mu\nu}\right]^{-1/2} \left[ \det p^2\delta^{ab}\right] \left[\det \mathcal{Q}^{ab}_{\alpha\beta}\right]^{1/2}\exp\left\{\gamma^4V4(N^2-1)\right\}\nonumber \;,\\
&=&  \exp \left\{\gamma^4V 4(N^2-1) -\frac{1}{2} \text{Tr} \ln \mathcal{P}^{ab}_{\mu\nu} +\text{Tr} \ln p^2\delta^{ab}+ \frac{1}{2} \text{Tr}\ln \mathcal{Q}^{ab}_{\alpha\beta}\right\}\;,
\end{eqnarray}
where
\begin{eqnarray}
\text{Tr}\ln \mathcal{P}^{ab}_{\mu\nu}& =& (N^{2} -1) V \int \frac{d^4 p}{(2\pi)^4} 
\left[ 4\ln p^2+
3 \ln \left(1 + 2N g^2 \gamma^4\frac{1}{p^4}\right)\right]\,,
\\
\text{Tr}\ln p^2\delta^{ab} &=& (N^2-1)V \int \frac{d^4 p}{(2\pi)^{4}} \ln p^2\,,
\\
\text{Tr}\ln \mathcal{Q}^{ab}_{\alpha\beta} &=& 2(N^2 -1) V \int \frac{d^4 p}{(2\pi)^{4}}
\left[\ln p^2+
 \ln\left(1+ \frac{M^6}{p^6}\right)\right]\;.
\end{eqnarray}
Therefore, the leading order result for the vacuum energy in the Gribov-Zwanziger framework is
\begin{eqnarray}
 \mathcal{E}_{v}(\gamma,M)&=& -\frac{1}{V} \ln Z_\text{quad} 
 \nonumber
\\
&=&-\gamma^4 4(N^2-1)+\frac{3}{2}(N^2 -1) \int \frac{d^4 p}{(2\pi)^4} \ln \left(1 + 2N g^2 \gamma^4\frac{1}{p^4}\right)\nonumber\\ 
&&- (N^2 -1) \int \frac{d^4 p}{(2\pi)^4} \ln \left(1+ \frac{M^6}{p^6}\right)\,,
\label{vaceq}
\end{eqnarray}
where the vanishing of the vacuum energy for $\mathcal{N}=1$ Super-Yang-Mills theories in the absence of the Gribov horizon, i.e. in the limit $\gamma,M\to 0$, is clear. This exact result to leading order stemms from the counting of bosonic and fermionic degrees of freedom, each contributing, respectively, with a negative and a positive $(V\int \frac{d^4 p}{(2\pi)^4}\ln p^2)$ term in the exponent of eq.\eqref{z1}. \\\\The next step is to apply the Gribov gap equation (cf. eq. \eqref{ggap}), 
\begin{equation}
\left. \frac{\partial \mathcal{E}_v(\gamma,M\bar\mu,g)}{\partial\gamma^2}\right|_{\gamma^*}=0\;,   \label{ggap1}
\end{equation}
in order to derive the nonperturbative expression for the Gribov parameter $\gamma$. From eq.\eqref{vaceq}, we have:
\begin{equation}
-4 + 3N g^2 \int \frac{d^4 p}{(2\pi)^4} \frac{1}{p^4 + 2N g^2\gamma^{*4}} =0 \;.
\end{equation}
Introducing the notation $\gamma'^4=2N g^2\gamma^{*4}$, one obtains
\begin{equation}
3N g^2 \int \frac{d^{4}p}{(2\pi)^4} \frac{1}{p^{4} + \gamma'^{4}} = 4\;,
\end{equation}
or, equivalently  (using $d=4-\epsilon$, in dimensional regularization),
\begin{equation}
N g^2(d-1)I_{\gamma}^d = d\,,
\end{equation}
with
\begin{eqnarray}
(d-1)I_{\gamma}^d&\equiv& (d-1)\frac{1}{2i\gamma'^2} \int \frac{d^dp}{(2\pi)^d}\left(\frac{1}{p^2-i\gamma'^2} - \frac{1}{p^2+i\gamma'^2}\right) 
\nonumber\\
&=&
(3-\epsilon)\frac{1}{(4\pi)^2}\left[\frac{2}{\epsilon}+1-\ln \left(\frac{\gamma'^2}{\bar\mu^2}\right)+O(\epsilon)\right]
\,.
\label{IgD}
\end{eqnarray}
where $\bar\mu$ is the  $\overline{\text{MS}}$ renormalization scale.
The gap equation then becomes, in the  $\overline{\text{MS}}$ renormalization scheme:
\begin{equation}
\label{gp1}
3N g^2\frac{1}{16\pi^2}\left[-\ln\frac{\gamma'^2}{\bar\mu^2}+ \frac13\right] = 4\;.
\end{equation}
The final non-perturbative expression for the Gribov parameter reads
\begin{equation}
\label{rslt}
\sqrt{2N}[\gamma^*(\bar\mu)]^2 = \frac{\bar\mu^2}{g}e^{ \left( \frac{1}{3} -\frac{16\pi^2}{3}\frac{4}{g^2 N}\right)} \;,
\end{equation}
Finally, using this result in eq.\eqref{vaceq}, the vacuum energy for $\mathcal{N}=1$ Super-Yang-Mills theories in the Gribov-Zwanziger framework becomes:
\begin{equation}
	\mathcal E_v = \frac{7(N^2-1)}{4(4\pi)^2} 
	2Ng^2[\gamma^*(\bar\mu)]^4
%	\bar\mu^4 e^{\frac{2}{3} -\frac{32\pi^2}{3}\frac{4}{g^2 N}} 
	- (N^2 -1) \int \frac{d^4 p}{(2\pi)^4} \ln \left(1+ \frac{M^6}{p^6}\right)  \;,
	\label{Ev-1}
\end{equation}
where we have used eq.\eqref{IgD} and the following relation:
\begin{eqnarray}
 \int \frac{d^4 p}{(2\pi)^4} \ln \left(1 + 2N g^2 \gamma^4\frac{1}{p^4}\right)
 =4Ng^2 \int_0^{\gamma^2} d(x^2) x^2 I_{x}^4\,.
\end{eqnarray}
The remaining momentum integral in eq.\eqref{Ev-1} is finite and can be solved directly,
\begin{equation}
\int \frac{d^4 p}{(2\pi)^4} \ln \left(1+ \frac{M^6}{p^6}\right) 
%=\frac{2\pi^2}{(2\pi)^4} \frac{\pi}{2\sqrt{3}}M^4
=\frac{1}{16\sqrt{3} \pi}M^4\,,
\end{equation}
so that the final expression for the vacuum energy as a function of the parameter $M$ reads:
\begin{equation}
	\mathcal E_v = \frac{7(N^2-1)}{4(4\pi)^2} 
	2Ng^2[\gamma^*(\bar\mu)]^4
%	\bar\mu^4 e^{\frac{2}{3} -\frac{32\pi^2}{3}\frac{4}{g^2 N}} 
	- (N^2 -1)\frac{1}{16\sqrt{3} \pi}M^4 \;,
	\label{Ev-1}
\end{equation}
It is now clear that there exists a nonzero solution $M^*(\bar\mu)$  which cancels out the vacuum energy produced by the presence of the Gribov parameter $\gamma$, recovering in this way the exact supersymmetric result $\mathcal E_v=0$. Furthermore, as discussed in the last section, a nonzero value of the mass parameter $M$ in the gluino propagator is also a necessary condition for reproducing a well-known feature of $\mathcal{N}=1$ Super-Yang-Mills theories: the gluino condensation (cf. eq.\eqref{gluinocondM}).
In view of these points, the modified gluino propagator may be regarded as a natural requirement of a consistent supersymmetric version of the Gribov-Zwanziger formalism. \\\\Explicitly, the vanishing of the vacuum energy gives the following expression for $M^*(\bar\mu,g)$, i.e. the supersymmetric counterpart of the Gribov parameter:
\begin{eqnarray}
\big[M^*(\bar\mu)\big]^4&=&
\frac{7\sqrt{3}}{4\pi}\, \gamma'^4
\;\;=\;\;
\frac{7\sqrt{3}}{4\pi}\, \bar\mu^4{\rm e}^{\frac{2}{3}-\frac{32\pi^2}{3}\frac{4}{g^2N}}
 \;, \label{zen}
\end{eqnarray}
whereas the gluino condensate, eq.\eqref{gluinocondM}, takes the form:
\begin{eqnarray}
\langle \bar\lambda \lambda \rangle_{cond}  &=& 
4(N^{2}-1) 
\left(\frac{7\sqrt{3}}{4\pi}\right)^{\frac{3}{4}}\, \gamma'^3
\sum_{i=1}^{3} \frac{\alpha_im_i^2}{(4\pi)^2} \left(\ln\frac{m_i^2}{\bar\mu^2} - 1 \right)   \nonumber
\\
&=&
4(N^{2}-1) 
\left(\frac{7\sqrt{3}}{4\pi}\right)^{\frac{3}{4}}
\, \bar\mu^3{\rm e}^{\frac{1}{2}-32\pi^2\frac{1}{g^2N}}
\sum_{i=1}^{3} \frac{\alpha_im_i^2}{(4\pi)^2} \left(\ln\frac{m_i^2}{\bar\mu^2} - 1 \right)\;.
\label{gluinocond}
\end{eqnarray} 
Collecting the results for the nonperturbative parameters that define the Gribov-Zwanziger extended supersymmetric Yang-Mills action,
\begin{eqnarray}
\big[\gamma^*(\bar\mu)\big]^4  & =& \frac{1}{2Ng^2} \bar\mu^4 {\rm e}^{ \left( \frac{2}{3} -\frac{32\pi^2}{3}\frac{4}{g^2 N}\right)} \;,
\\
\big[M^*(\bar\mu)\big]^4&=&
\frac{7\sqrt{3}}{4\pi}\, \gamma'^4
\;\;=\;\;
\frac{7\sqrt{3}}{4\pi}\, \bar\mu^4{\rm e}^{\left(\frac{2}{3}-\frac{32\pi^2}{3}\frac{4}{g^2N}\right)}
 \;,
\end{eqnarray} 
inspection reveals that they are both related to a single nonperturbative physical scale, currently encoded in the $\overline{\text{MS}}$ renormalization scale. This is consistent with the fact the theory has only one physical scale,
 in the same way as massless QCD presents only the confinement scale $\Lambda_{QCD}$.
The expressions above may also be recast in a renormalization-scheme independent form, in terms of this nonperturbative physical (i.e. renormalization-group invariant) scale. Making use of the definition of such a quantity,
\begin{eqnarray}
\left(\bar\mu\frac{\partial}{\partial\bar \mu}+\beta_g\frac{\partial}{\partial g}\right)\Lambda^{N=1}_{SYM}&=&0\,,
\end{eqnarray}
with $\beta_g\equiv\bar\mu\frac{\partial}{\partial\bar \mu}g$,
and of the one-loop $\beta$ function of $\mathcal{N}=1$ Super-Yang-Mills theories, $\beta_g=-\beta_0 \frac{g^3}{(4\pi)^2}$ (with $\beta_0=3N$, see for instance \cite{Mihaila:2013wma}), one obtains:
\begin{eqnarray}
\Lambda^{N=1}_{SYM} &=&\bar\mu {\rm e}^{-\frac{(4\pi)^2}{2}\frac{1}{\beta_0g^2}}
\end{eqnarray}
and
\begin{eqnarray}
\big[\gamma^*\big]^4 &=& \frac{1}{2Ng^2} (\Lambda^{N=1}_{SYM})^4 {\rm e}^{ \left( \frac{2}{3} -32\pi^2\frac{1}{g^2 N}\right)} \;, \nonumber 
\\
\big[M^*\big]^4&=&
\frac{7\sqrt{3}}{4\pi}\, (\Lambda^{N=1}_{SYM})^4 {\rm e}^{\left(\frac{2}{3}-32\pi^2\frac{1}{g^2N}\right)}\,.   \label{npert}
\end{eqnarray}

%%%%%%%%%%%%%%%%%%%%
\section{Conclusion}
%%%%%%%%%%%%%%%%%%%%
In this work we have presented the extension of the Gribov-Zwanziger framework to $\mathcal{N}=1$ Super-Yang-Mills theories quantized in the Wess-Zumino gauge, by imposing the Landau gauge condition. Our construction is summarized by the action $S_{SGZ}^{N=1} $, given in expression  \eqref{lcsgzn1}. This action has the meaning of an effective action encoding the  restriction to the first Gribov horizon in a way compatible with non-perturbative supersymmetric features. This has been possible due to the presence in expression \eqref{lcsgzn1} of two massive parameters $(\gamma, M)$, which have been obtained in a dynamical way through suitable non-perturbative conditions. \\\\The parameter $\gamma$,  determined by the gap equation \eqref{ggap1}, is the Gribov parameter which arises as the consequence of the restriction of the domain of integration in the Euclidean path integral to the Gribov region $\Omega$. The second parameter $M$ can be regarded as a kind of supersymmetric counterpart of the Gribov parameter $\gamma$. Its presence is needed in order to consistently ensure the vanishing of the vacuum energy, eqs.\eqref{vc},\eqref{zen}, as required by supersymmetry. The two conditions \eqref{ggap1},\eqref{vc} enable us to determine the two parameters $(\gamma, M)$ in a non-perturbative way, as expressed by \eqref{npert}. Moreover, in agreement with supersymmetry, a non-vanishing gluino condensate is found, eq.\eqref{gluinocond}. \\\\Besides recovering non-perturbative features of supersymmetry, the action   \eqref{lcsgzn1} is suitable to study the confinement of the elementary degrees of freedom, {\it i.e.} of gluon and gluinos, as one can infer from the presence of complex poles in the corresponding two-point correlation functions, eqs.\eqref{gg}, \eqref{lqu1}, which can be seen as a strong indication of the absence of these excitations from the physical spectrum. \\\\To some extent, the action $S_{SGZ}^{N=1} $ represents the first step in order to address the issue of the Gribov copies in supersymmetric gauge theories, opening the possibility of investigating within a local and renormalizable field theory other non-perturbative aspects, such as: supersymmetric generalization of the so called Refined-Gribov-Zwanziger framework, estimates of the masses of the low lying states of the spectrum and comparison with the available numerical lattice data, extension to $\mathcal{N}=2$ Super Yang-Mills theories and study of the corresponding phase diagram, analysis of the Gribov issue in $\mathcal{N}=4$ theories and investigation of a possible relationship with Maldacena's conjecture. We hope to report soon on these interesting topics.

%%%%%%%%%%%%%%%%%%%%%%%%%%%
\section*{Acknowledgments}
%%%%%%%%%%%%%%%%%%%%%%%%%%%
The Conselho Nacional de Desenvolvimento Cient\'{\i}fico e
Tecnol\'{o}gico (CNPq-Brazil), the Faperj, Funda{\c{c}}{\~{a}}o de
Amparo {\`{a}} Pesquisa do Estado do Rio de Janeiro,  the
Coordena{\c{c}}{\~{a}}o de Aperfei{\c{c}}oamento de Pessoal de
N{\'{\i}}vel Superior (CAPES)  are gratefully acknowledged.

\end{document}